\documentclass{llncs}
\usepackage{llncsdoc}
\usepackage{epsfig}
\usepackage{graphicx}
\usepackage[leqno]{amsmath}
\usepackage{amssymb}
\usepackage{mathrsfs}
\usepackage{psfrag}
\usepackage[hyphenbreaks]{breakurl}
\usepackage[hyphens]{url}
\usepackage[usenames]{color}

\usepackage{braket} % per il comando \Set: definizione automatica di insiemi
\usepackage{subfig}
\usepackage{varioref}       % per indicazione di pagina nei riferimenti incrociati
\usepackage{enumerate}
\usepackage{cleveref}

\crefname{figure}{Fig.}{Figs.}
\newcommand{\Z}{\mathbb{Z}}
\newcommand{\R}{\mathbb{R}}

\newcommand{\T}{\mathbb{T}}

\newcommand{\rvec}[1]{\left( #1 \right)} % vettore

\renewcommand{\leq}{\leqslant}

\begin{document}

\title{Towards a topological fingerprint of music}
\titlerunning{Towards a topological fingerprint of music}  % abbreviated title (for running head)
\author{Mattia G. Bergomi\inst{1},  Adriano Barat\`{e}\inst{2} \and Barbara Di Fabio\inst{3}}
\authorrunning{M. G. Bergomi \and A. Barat\'{e} \and B. di Fabio} % abbreviated author list (for running head)
%Craig Chambers, Kim B. Bruce, and Elisa Bertino}
%
\institute{Champalimaud Neuroscience Programme, Champalimaud Centre for the Unknown, Lisbon, Portugal,\\
\email{mattia.bergomi@neuro.fchampalimaud.org} \and Universit\`{a} degli Studi
di Milano, Laboratorio di Informatica Musicale\\Milano, Italy\\
\email{barate@di.unimi.it} \and Universit\`{a} di Modena e Reggio
Emilia, Dipartimento di Scienze
e Metodi dell'Ingegneria\\ Reggio Emilia, Italy\\
\email{barbara.difabio@unimore.it}}

\maketitle              % typeset the title of the contribution

\begin{abstract}
Can music be represented as a meaningful geometric and topological
object? In this paper, we propose a strategy to describe some
music features as a polyhedral surface obtained by a simplicial
interpretation of the \textit{Tonnetz}. The \textit{Tonnetz} is a graph largely used
in computational musicology to describe the harmonic relationships of
notes in equal tuning. In particular, we use persistent homology
in order to describe the \textit{persistent} properties of music
encoded in the aforementioned model. Both the relevance and the
characteristics of this approach are discussed by analyzing some
paradigmatic compositional styles. Eventually, the task of
automatic music style classification is addressed by computing the
hierarchical clustering of the topological fingerprints associated
with some collections of compositions. 
\end{abstract}
\section{\label{sec:introduction} Introduction}
Generally, the core of a piece of music consists of a small
collection of strong, recognizable concepts, that are
grasped by the majority of the
listeners~\cite{dowling1972recognition,folgieri2014eeg,lara2015}.
These \textit{core concepts} are developed during the composition by varying levels of
tension over time, drawing the attention of the listener to
particular moments thanks to specific choices, frustrating his/her
intuition through unexpected changes, or confirming his/her
expectation with, for instance, a well-known cadence leading to
resolution.

As the models for the analysis of audio signals take advantage
of the strategies developed for image
analysis~\cite{smaragdis2003non,wang2003industrial,li2010automatic},
it is possible to borrow some tools from the topological analysis
of shapes and data to tackle the problem of music analysis and
classification. The main aim of this paper is the introduction of
a low-dimensional geometric-topological  model in order to
describe, albeit in an extremely simplified form, music styles.

Loosely speaking, we introduce a metric representation of music as a
planar polyhedral surface, whose vertices are then translated
along a third dimension in basis on a specific function. The
shapes obtained via these deformations are fingerprinted by
computing their \textit{persistent
homology}~\cite{edelsbrunner2008persistent}. Afterwards, the
musical meaning of this topological representation of music is
discussed and applied to automatic style classification on three
different datasets.

%\Cref{sec:tonnetz} provides the definitions of the main musical
%ingredients that shall be used in this paper and gives an insight
%on the problem of the topological characterization of musical
%entities. In~\Cref{sec:anis} a novel method to encode part of the
%music information in a $3$-dimensional shape is presented.
%Persistent homology is introduced in~\Cref{sec:fing}, together
%with several examples and applications.

\section{Background on persistent homology}

In computational topology \cite{EdHa09}, persistent homology is
actually considered an invaluable tool to describe both geometry
and topology of a certain space, not only because of the
simplicity of the method, but also because all the properties are
ranked by importance, allowing us to choose the level of detail at
which to perform such a
description~\cite{edelsbrunner2008persistent,FrLa99}.

\begin{figure}[tb]
\centering
\includegraphics[width=\textwidth]{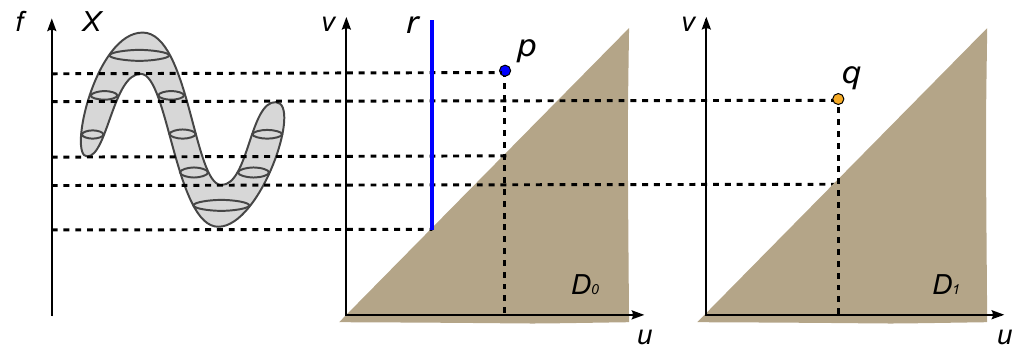}
\caption[persistence diagrams.]{Left: The height function on the
topological space $X$. Right: The associated persistence diagrams
$D_k(X,f)$, with $k=0,1$. \label{fig:persEx}}
\end{figure}

In more formal terms, given a topological space $X$, we define a
continuous function $f:X\to \R$ to obtain a family of subspaces
$X_u=f^{-1}((-\infty,u]), u\in\R$, nested by inclusion, i.e. a
filtration of $X$. The map $f$, called therefore a \emph{filtering
function}, is chosen according to the geometrical properties of
interest (e.g., height, distance from center of mass, curvature).
Applying homology to the filtration, births and deaths of
topological features can be algebraically detected and their
lifetime measured. The scale at which a feature is significant is
measured by its longevity. Formally, given $u \le v \in\R$, we
consider the inclusion of $X_u$ into $X_v$. This inclusion induces
a homomorphism of homology groups $H_k(X_u)\to H_k(X_v)$ for every
$k \in\Z$. The image of such a homomorphism consists of the
$k$-homology classes that live at least from $u$ to $v$ along the
filtration, and is called the $k$th \emph{persistent homology
group} of the pair $(X,f)$ at $(u,v)$. When, for every $(u,v),
u\le v$, the $k$th persistent homology groups are finitely
generated, we can compactly describe them using the so-called
\emph{persistence diagrams}. A persistence diagram $D_k(X,f)$ is
the subset of $\{(u,v)\in\R^2: u< v\}$ consisting of points
(called \emph{proper points}) and vertical lines (called
\emph{points at infinity}) encoding the levels of $f$ at which
the birth or the death of homological classes occur, union all the points
belonging to the diagonal $u=v$. In particular, if there exists at
least one $k$-homology class that is born at the level $\bar u$
and is dead at the level $\bar v$ along the filtration induced by
$f$, then $p=(\bar u,\bar v)$ is a proper point of $D_k(X,f)$; if
there exists at least one $k$-homology class that is born at the
level $\bar u$ and never dies along the filtration induced by $f$,
then $p=(\bar u,+\infty)$ is a point at infinity of $D_k(X,f)$. A
point at infinity is usually represented as the vertical line
$u=\bar u$. Both points and lines are equipped with a multiplicity
that depends on the number of classes with the same
lifetime~\cite{FrLa01}. An example of persistence diagrams is
displayed in~ \Cref{fig:persEx}. The surface $X\subset\R^3$ is
endowed with the height function $f$. The associated persistence
diagrams $D_0(X,f)$ and $D_1(X,f)$ are displayed on the right. $D_0(X,f)$ consists of one point at infinity $r$, whose abscissa $u$
detects the absolute minimum of $f$, and one proper point $p$,
whose abscissa and ordinate detect, respectively, the level at
which the new connected component appears and merges with the
existing one. $D_1(X,f)$ consists of one proper point, whose
abscissa and ordinate detect, respectively, the level at which a
new tunnel is created and disappears along the filtration.

One of the main reasons behind the usage of persistence diagrams
in applications consists in the possibility of estimating the
degree of dissimilarity of two spaces with respect to a certain
geometrical property through an appropriate comparison of these
shape descriptors. Because of its properties of
optimality~\cite{dAFrLa} and stability~\cite{CoEdHa07}, the most
used instrument to compare persistence diagrams is
given by the so called \emph{bottleneck distance} (a.k.a.
\emph{matching distance})~\cite{dAFrLabis}.

\begin{definition}
The bottleneck distance between two persistence diagrams $D$ and
$D'$ is defined as
\begin{equation*}
d_B(D,D')=\min_{\sigma}\max_{p\in D}d(p,\sigma(p)),
\end{equation*}
where $\sigma$ varies among all the bijections between $D$ and
$D'$ and
\begin{equation*}
d\left(\left(u,v\right),\left(u',v'\right)\right)=\min\left\{\max\left\{|u-u'|,|v-v'|\right\},\max\left\{\frac{v-u}{2},\frac{v'-u'}{2}\right\}\right\}
\end{equation*}
for every $\left(u,v\right),\left(u',v'\right)\in\{(x,y)\in\R^2:
x\le y\}$.
\end{definition}

\begin{figure}[tb]
\centering \subfloat[Pitch
space.]{\includegraphics[width=0.4\textwidth]{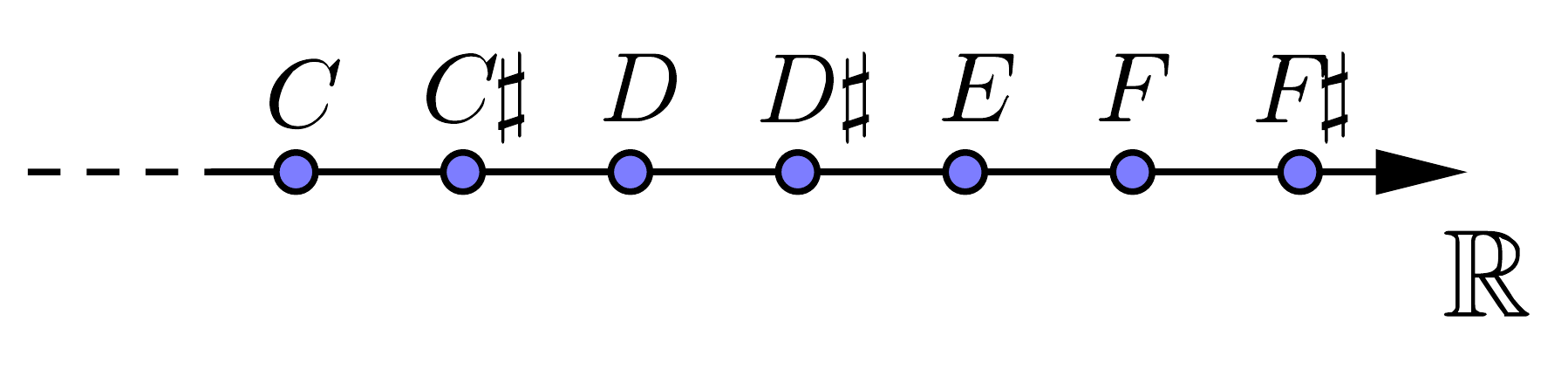}}\qquad\qquad\qquad
\subfloat[Pitch-class
space.]{\includegraphics[width=0.4\textwidth]{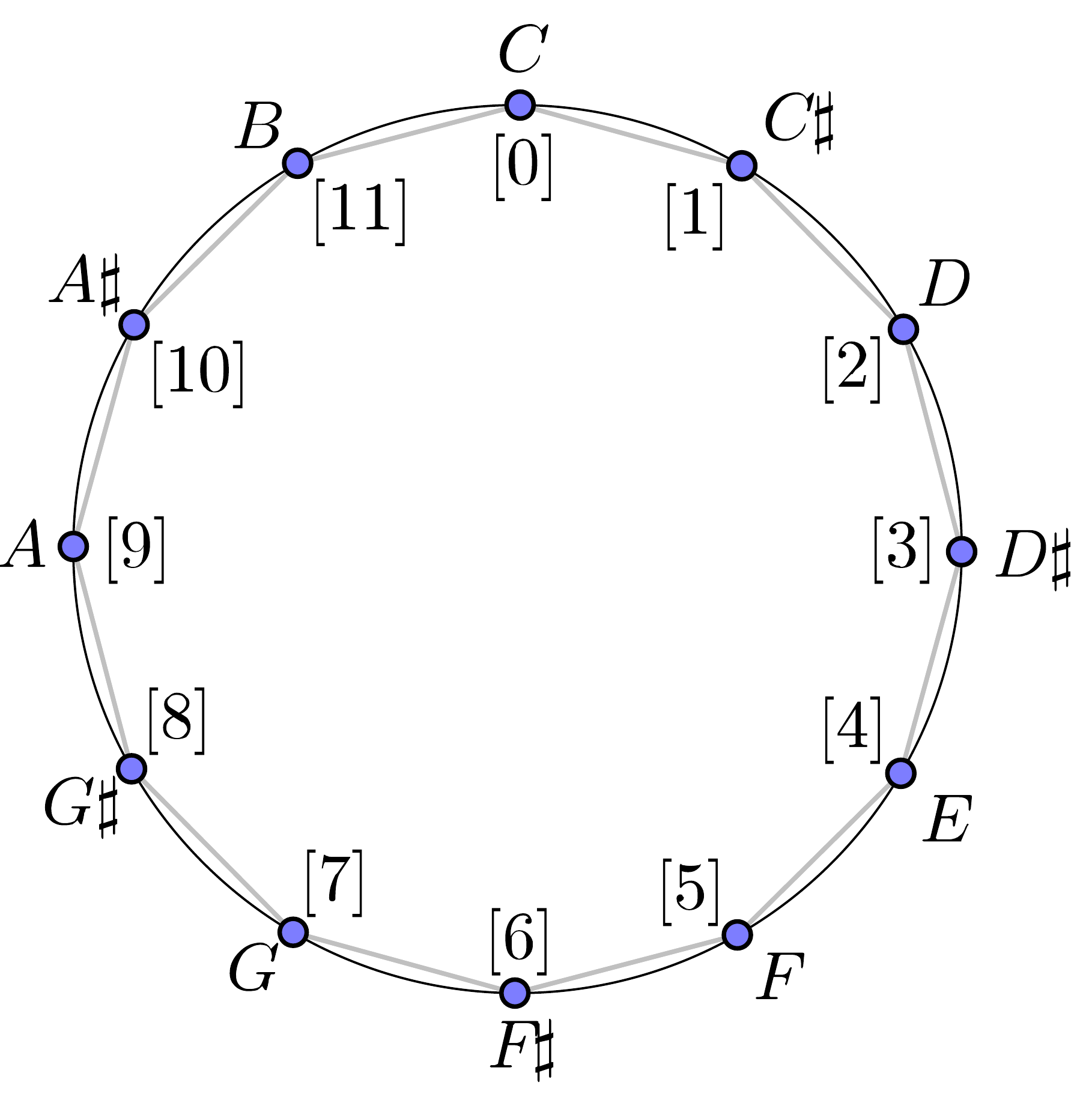}}
\caption{\label{fig:pitchlasses} Fundamental music representation
spaces.}
\end{figure}

\section{Musical setting}

In order to safely introduce the main model presented in this
paper, we start by defining some basic musical objects.

We model a \emph{note in equal tuning} $\mathfrak{n}$ as a pair
$(p,d)\in\R^2$, where $p$ is called the \emph{pitch} of the note,
and $d$ is its \emph{duration} in seconds. In particular, if $\nu$
denotes the fundamental frequency of $\mathfrak{n}$, the pitch $p(\nu)$ is
defined as $p(\nu) = 69 + 12\log_2 \left(\frac{\nu}{440} \right)$,
where $440Hz$ is the fundamental frequency of the note $A_4$ (the
\textit{la} of the fourth octave of the piano). For further
details on pitches, see, e.g.,~\cite{de2005pitch}.

On a perceptual level, two notes an octave apart are really
similar~\cite{burns1999intervals}, thus, it is common to identify
pitches modulo octave, by considering \emph{pitch classes} $[p]=
\Set{p+12k : k \in \Z}\cong\R/12\Z$. A representation of both the
pitch and pitch-class spaces is depicted
in~\cref{fig:pitchlasses}.

\subsection{\label{sec:tonnetz}The simplicial \textit{Tonnetz}}

\begin{figure}[tb]
\centering
\subfloat[\label{fig:torgen1}Planar simplicial\textit{Tonnetz}.]{\includegraphics[width=0.45\textwidth]{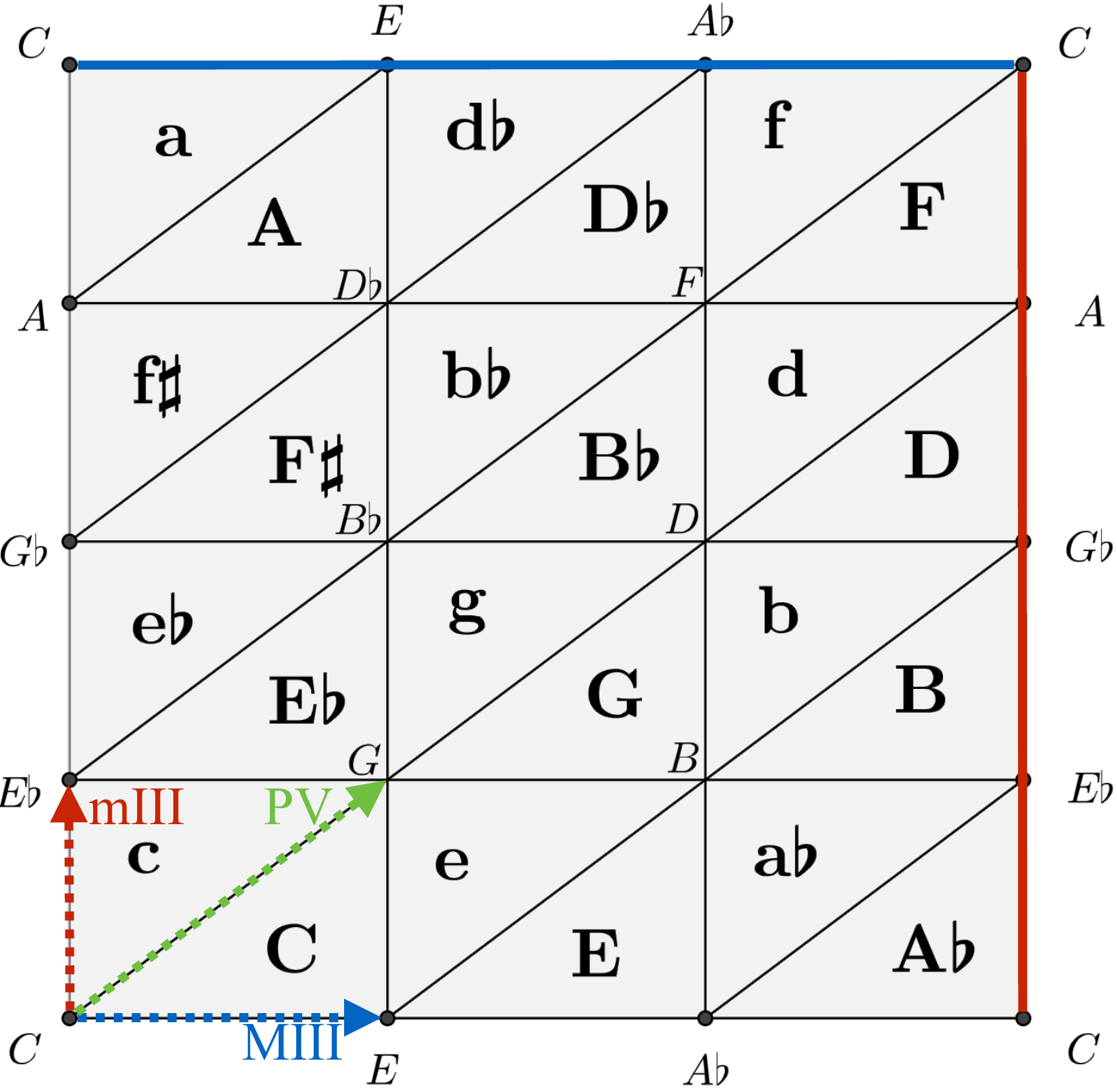}}\qquad
\subfloat[\label{fig:torgen2}\textit{Tonnetz} torus.]{\includegraphics[width=0.45\textwidth]{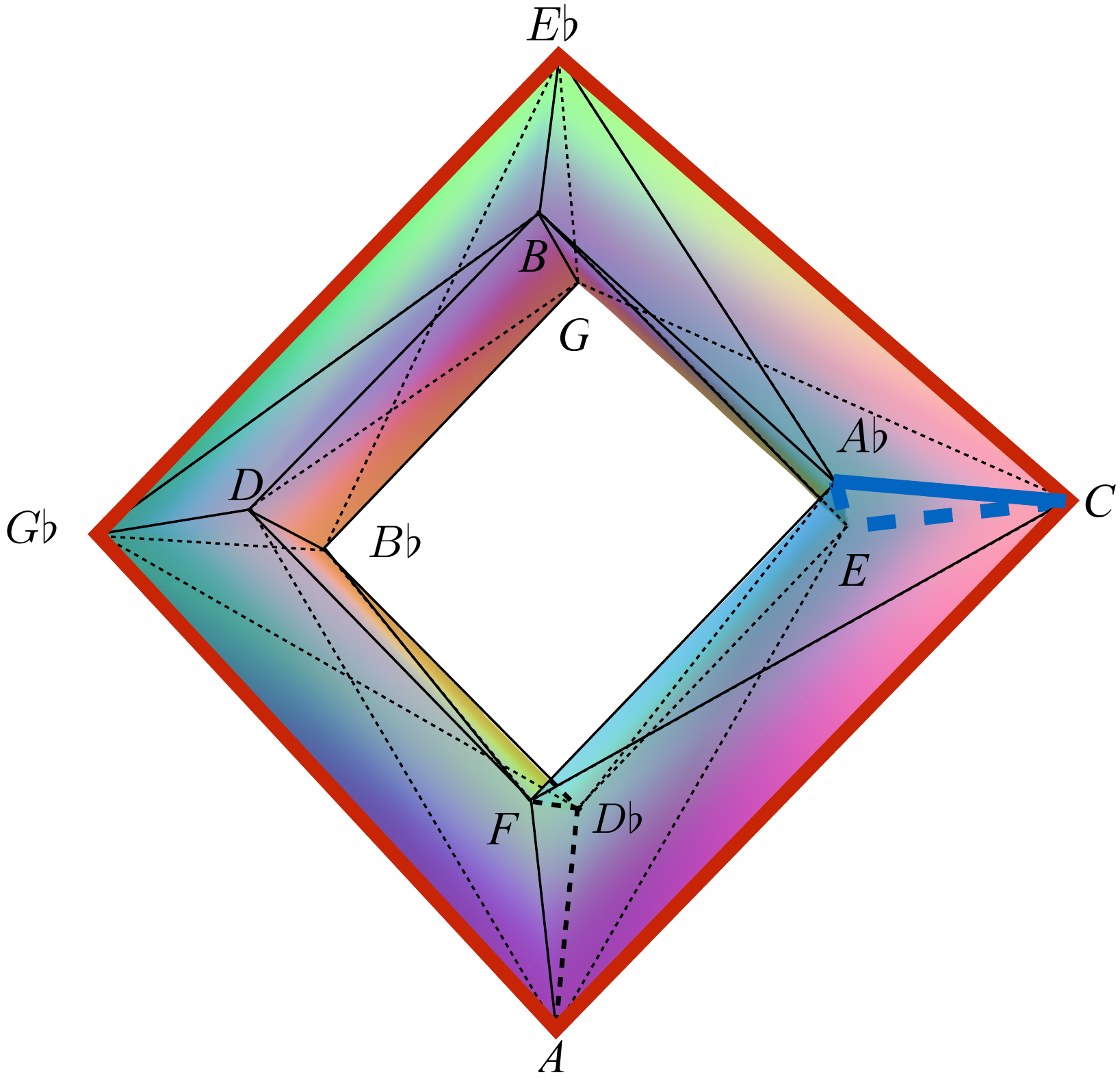}}
\caption{\label{fig:torgen} (a) A finite subcomplex of the
\textit{Tonnetz}. (b) The \textit{Tonnetz} torus $\T$ obtained by
identifying vertices in (a) equipped with the same labels.}
\end{figure}

The \emph{Tonnetz} was originally introduced in
\cite{euler1774harmoniae} as a simple $3\times 4$ matrix
representing both the acoustical and harmonic relationships among
 pitch classes. Later, it has been largely generalized to
several formalisms, see,
e.g.,~\cite{cohn1997neo,zabka2009generalized,douthett1998parsimonious}.
We will focus on its interpretation as a simplicial
complex~\cite{bigo2013computation}. In this setting, the
\textit{Tonnetz} is modeled as an infinite planar simplicial
complex, whose $0$-simplices are labeled with pitch classes in a
way that $1$-simplices form either perfect fifth, major, or minor
third intervals, and $2$-simplices correspond to either major or
minor triads. A finite subcomplex of the \textit{Tonnetz} $T$ is depicted
in~\cref{fig:torgen1}. We observe that the labels on its vertices
are periodic with respect to the transposition of both minor and
major third. This feature allows to work with the more
comfortable toroidal representation $\T$ displayed
in~\cref{fig:torgen2}.
\begin{figure}[tb]
\centering
\subfloat[The lydian mode.\label{fig:subcomp4}]{\includegraphics[width=\textwidth]{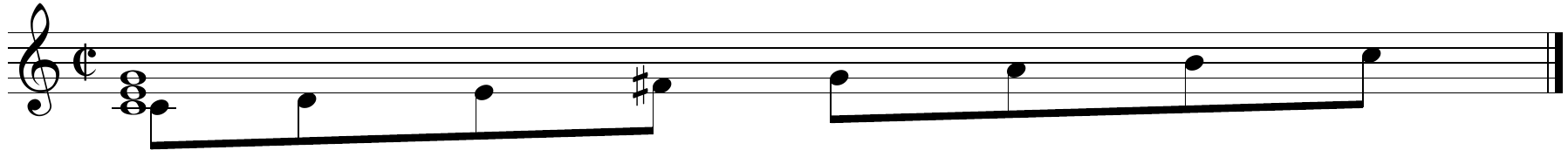}}\\[-3pt]

\subfloat[The locrian mode.\label{fig:subcomp5}]{\includegraphics[width=\textwidth]{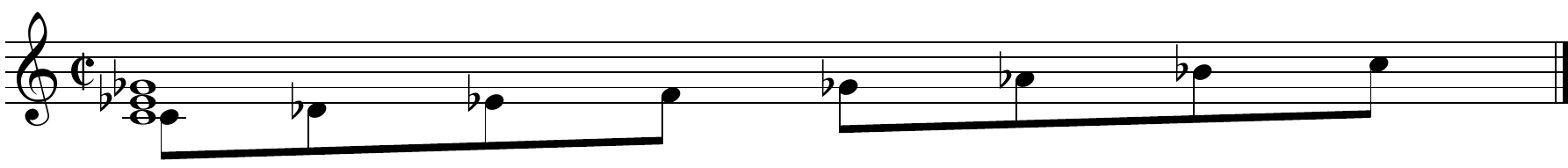}}\\[-2pt]

\subfloat[Ionian subcomplex.\label{fig:subcomp2}]{\includegraphics[width=0.47\textwidth]{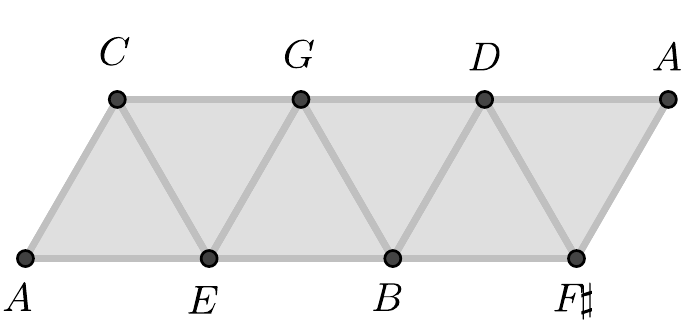}}\qquad
\subfloat[Locrian subcomplex.\label{fig:subcomp3}]{\includegraphics[width=0.45\textwidth]{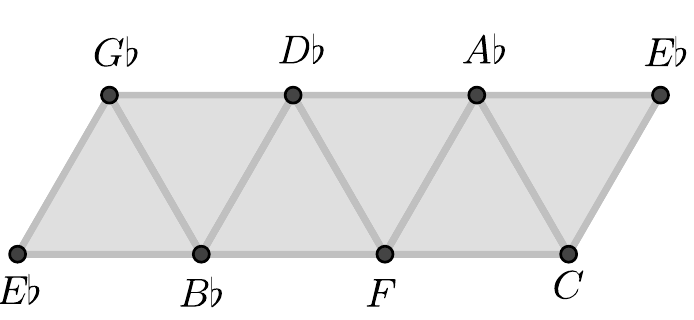}}

\caption{Two different modes  represented by isomorphic subcomplexes of the \textit{Tonnetz}.\label{fig:subcomp}}
\end{figure}

It is possible to analyze and classify music
by considering the subcomplexes of $T$ generated by a
sequence of pitch classes~\cite{bigo2013computation}. However,
this approach does not allow to discriminate
musical styles in a geometric or
topological sense. In fact, as the example in~\cref{fig:subcomp} shows, two perceptively distinct sonorities can be
represented by isomorphic subcomplexes.

\subsection{A deformed \textit{Tonnetz} for music
analysis}\label{sec:anis}

In order to capture both the \textit{temporal} and harmonic
information encoded in a musical phrase, the vertices of the
\textit{Tonnetz} shall be displaced depending both on the pitch which
is played and its duration.

\begin{figure}[t]
\begin{centering}
\subfloat[$\mathcal{T}$]{\begin{centering}
\includegraphics[width=0.4\textwidth]{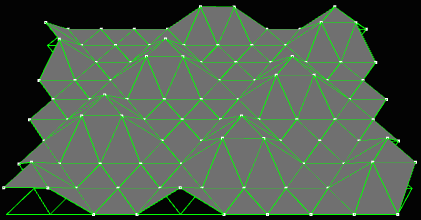}
\par\end{centering}
}
\hspace{0.5cm}
\subfloat[$\mathcal{T}^{(1)}$]{\begin{centering}
\includegraphics[width=0.4\textwidth]{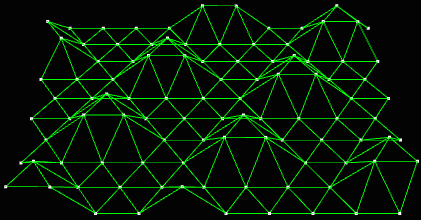}
\par\end{centering}
}
\par\end{centering}
\protect\caption[Visualization of the \textit{Tonnetz} simplicial
structure.]{The \textit{Tonnetz} deformed with a major triad (a)
and its $1$-skeleton (b). The triad appears as a maximal triangle
with respect to the height function.\label{fig:deftriad}}
\end{figure}

Let $V$ be the $0$-skeleton of $T$ and
$\{\mathfrak{n}_1,\dots,\mathfrak{n}_m\}=\{(p_1,d_1),\dots, (p_m,d_m)\}$ a finite
collection of notes of a musical phrase. Assume that
$\{\mathfrak{n}_{i_1},\dots,\mathfrak{n}_{i_k}\}$ is the subset whose pitches
$p_{i_1},\dots,p_{i_k}$ belong to $[p]$. We define a map that
takes each vertex $v=(x_v,y_v,0)\in \R^3$ labeled with $[p]$ to the
point $(x_v,y_v,d_v)\in\R^3$, where $d_v=\sum_{j=1}^k d_{i_j}$,
and then extend it  linearly to all the simplices. The
\textit{Tonnetz} deformed under the action of this map will be
denoted by $\mathcal{T}$, and will be used as the main object of our
topological description of music style. An example of deformation
induced by a major triad played for $8$ seconds is depicted
in~\Cref{fig:deftriad}, while a $3$-dimensional interactive animation showing how the
\textit{Tonnetz} is deformed by a musical phrase is available at \burl{http://nami-lab.com/tonnetz/examples/deformed_tonnetz_int_sound_pers.html}.

\section{A topological fingerprint of music
styles}\label{sec:fing}
In order to describe the deformed \textit{Tonnetz}, we use persistent homology.

We define the height
function $f$ on $\mathcal{T}$ to induce a  lower level set
filtration on the torus $\T$.
The persistence diagrams obtained with this process are
\textit{descriptors} of the style characterizing
the composition represented as a shape.
\begin{figure}[tb]
\centering
\includegraphics[width=\textwidth]{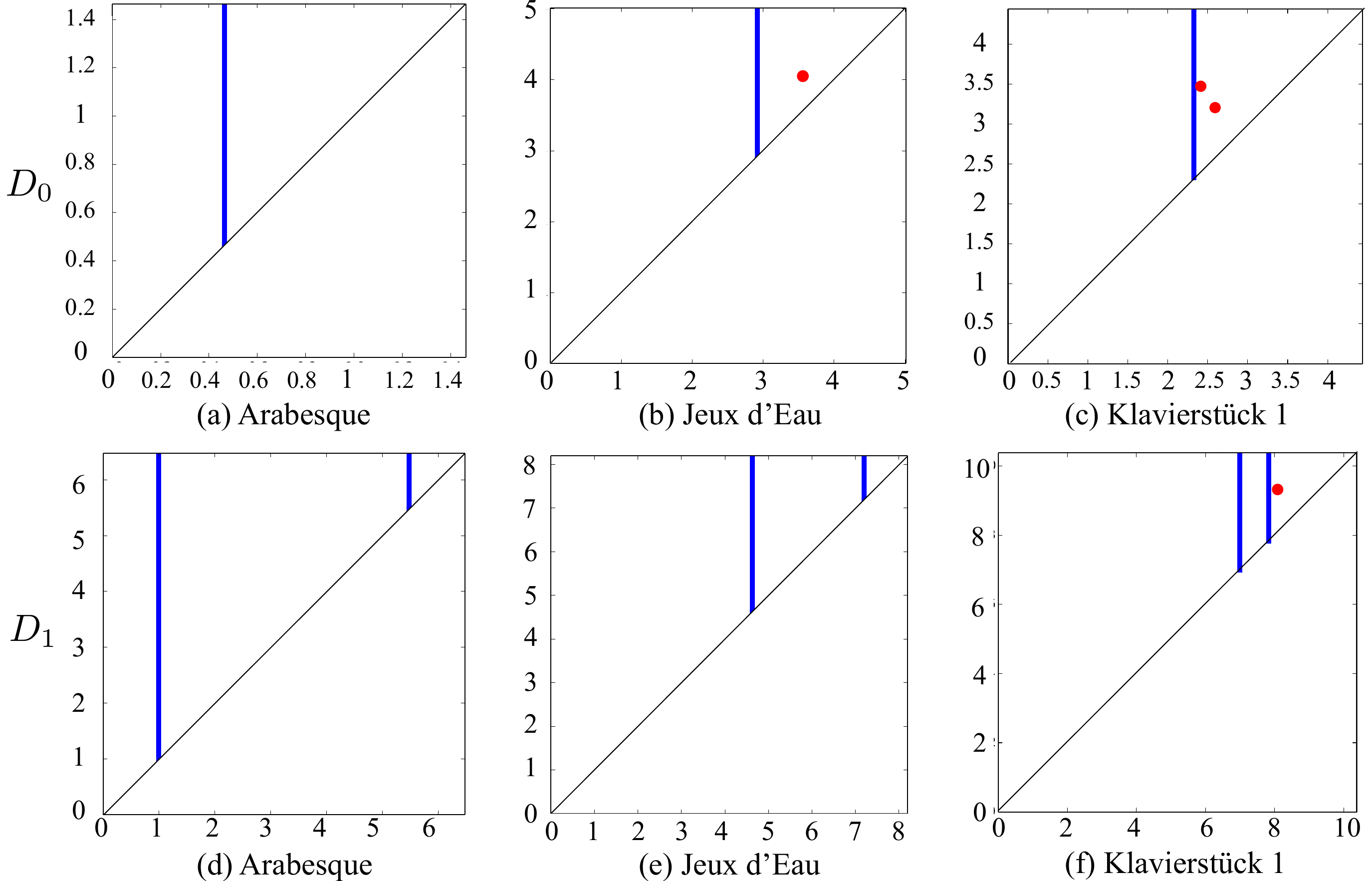}
\protect\caption{The $0$ (first row) and $1$st persistence diagrams (second row) representing
the topological fingerprints associated with three different
compositions.\label{fig:crit2}}
\end{figure}

\textbf{$0$th persistence diagrams.} The connectedness of $\T$ is
retrieved by the presence of only one point at infinity. Let $u =
\bar u$ be its equation: $\bar u$ is the absolute minimum of $f$
on the deformed \textit{Tonnetz}. If $\bar u\approx 0$, then there
exists at least one pitch-class set that does not have a relevant role
in the composition, suggesting that it is based on a stable tonal
or modal choice. On the contrary, if $\bar u>>0$, then all the
pitch classes have been used in the composition for a relevant
time. This configuration suits a more atonal or chromatic style. The presence of proper points is due to the existence of minima of the height function, that are subcomplexes of the \textit{Tonnetz} not
connected by an edge, and hence, representing a dissonant interval~\cite{plomp1965tonal}. Furthermore, the
structure of the \textit{Tonnetz} torus allows to retrieve a maximum of three connected components.
To create this particular configuration, it is necessary to play a
chromatic cluster: for instance, $C,C\sharp,D$, that is not
usually used in a tonal or modal context.

\textbf{$1$st persistence diagrams.} The lifespan of $1$-dimensional
holes traversing the filtration provides symmetrical information
with respect to the $0$th persistence analysis. In this case, two points at infinity detect the two generators of the $1$st homology group of the torus and, if there exists, proper points detect the presence of maxima of the height function, that are subcomplexes of the \textit{Tonnetz} not
connected by an edge.

As an example, we consider the persistence diagrams associated with
 Debussy's \textit{Arabesque}, \textit{Jeux d'Eau} by
Ravel, and \textit{Klavierst\"{u}ck 1} by Sch\"{o}nberg, shown in~\cref{fig:crit2}. In the $0$th persistence diagram
describing \textit{Arabesque}, there are no proper points. This is
an evidence of the pentatonic and diatonic/modal
inspiration of the composition~\cite{trezise2003cambridge}. We also observe that the entire chromatic scale has been used,
since $\bar u>0$. The abscissa of the point at infinity in the $0$th
persistence diagram of \textit{Jeux d'Eau} is
characterized by a high value, thus the entire set of pitch classes has been largely used in the composition. Moreover, the presence of a proper point highlights
 the \textit{ante-litteram} use of the \textit{Petrushka chord},
a superposition of a major triad and its tritone substitute: for
instance, $G = \rvec{G,B,D} + C\sharp
=\rvec{C\sharp,E\sharp,G\sharp}$. Finally, the diagram associated
with the \textit{Klavierst\"{u}ck 1} has two relevant proper points: this last feature
points out the atonal nature of the composition.

The second row of~\Cref{fig:crit2} shows the $1$st persistence diagrams associated with the same compositions. The tonal nature of \textit{Arabesque} is highlighted by the a large distance between the points at infinity and the absence of proper points. The chromatic style of \textit{Jeux d'Eau} implies the reduction of the distance between the two points at infinity. This last feature appears also in the diagram describing the \textit{Klavierst\"{u}ck 1}, whose atonal tendency is stressed by a proper point representing the relevant lifespan of a third non-connected subcomplex.

\begin{figure}[tb]
\begin{centering}
\includegraphics[width=\textwidth]{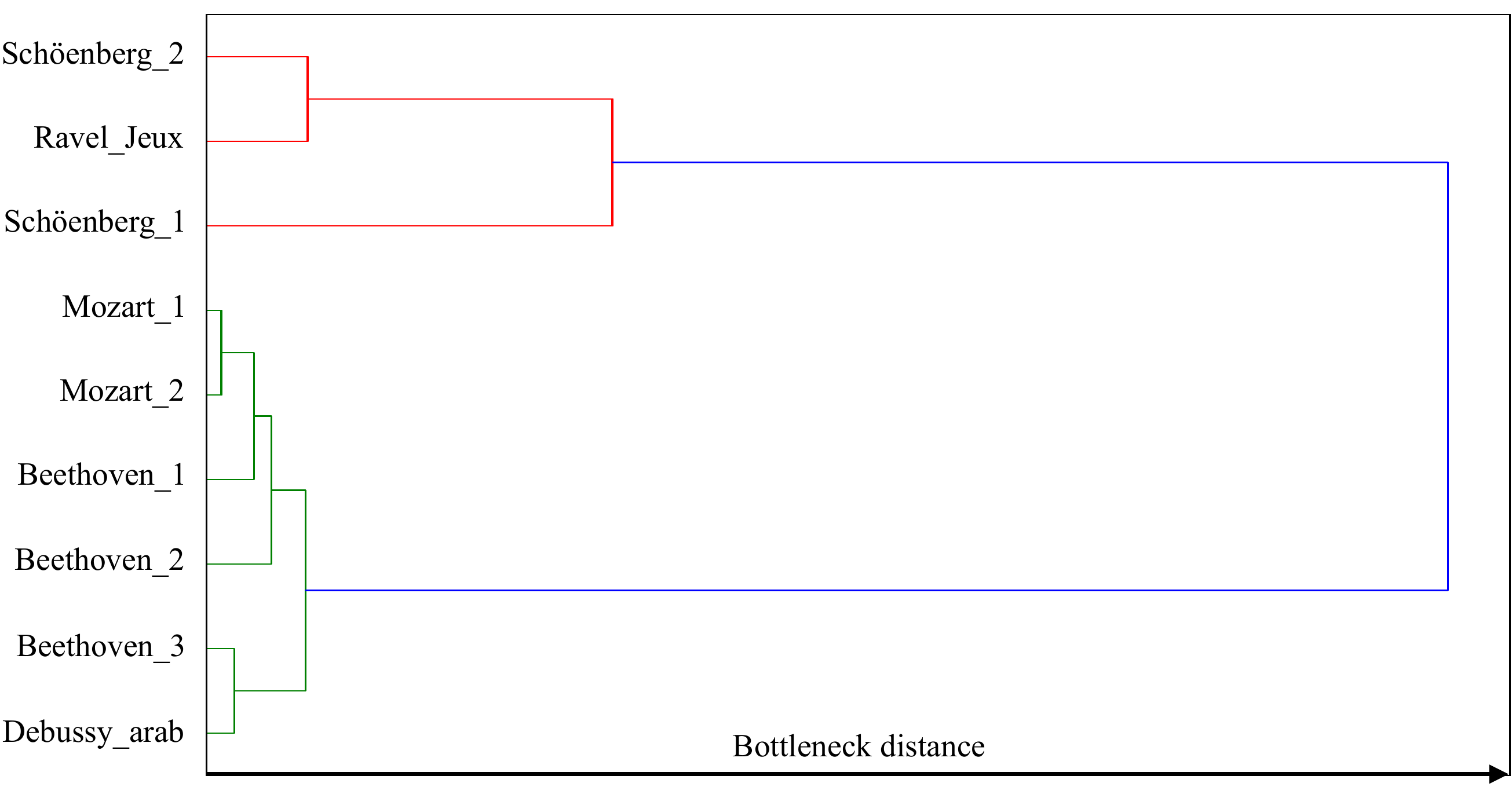}
\par\end{centering}
\protect\caption{Persistence-based clustering of nine classical and contemporary pieces.\label{fig:dend0cla}}
\end{figure}

\subsection{Applications}

In the following applications, we show how the persistence diagrams associated with a collection of compositions can effectively classify them according to their style. For $k=0,1$, let $\mathcal{D}= \{D_1,\dots, D_n\}$ be the set of $k$th persistence diagrams. Let $M = m_{ij}=d_B(D_i,D_j)$, for $1\leq i,j\leq n$, be their distance matrix.
The hierarchical
clustering analysis~\cite{ott2009visualization} allows us to describe
the configuration of the diagrams  $D_i\in\mathcal{D}$ with respect to the
bottleneck distance. We will represent the
organization of all their possible clusters  as a
dendrogram~\cite{langfelder2008defining,martinez2010exploratory}. In this type of diagram, the abscissa of each splitting (vertical line) measures the distance between two clusters. Such distance is computed through elementary operations on the elements of $M$.
\newline\newline
\textbf{Tonal and Atonal Music.}
We consider a dataset composed by nine pieces selected among the
compositions by Beethoven, Debussy,  Mozart, Ravel and
Sch\"{o}nberg available at
\url{http://nami-lab.com/tonnetz/examples/deformed_tonnetz_int_sound_pers.html}.

The clustering computed using the $0$th persistence of these pieces is depicted
in~\cref{fig:dend0cla}. Data are organized in two main clusters, that segregate the two
first pieces of Sch\"{o}nberg's \emph{Drei Klavierst\"{u}cke} and
Ravel's \emph{Jeux d'Eau}, from the ones by Mozart, Beethoven and
Debussy. The association between
\emph{Klavierst\"{u}ck 2} and \emph{Jeux d'Eau} mirrors the
particular nature of this Sch\"{o}nberg's composition, that lies
at the crossroad of tonal and atonal music, as it is proven by its
disparate tonal
interpretations~\cite{brinkmann2000arnold,william1984harmony,ogdon1981tonality}. The two movements of Mozart's \textit{KV311} form immediately a cluster
reached at an increasing distance by the two first movements of the
\textit{Sonata in C major} by Beethoven. The third movement of \textit{Sonata in C major}  is
grouped with \emph{Arabesque} because both are characterized
by a generous use of the pentatonic scale.
\begin{figure}[tb]
\begin{centering}
\includegraphics[width=0.9\textwidth]{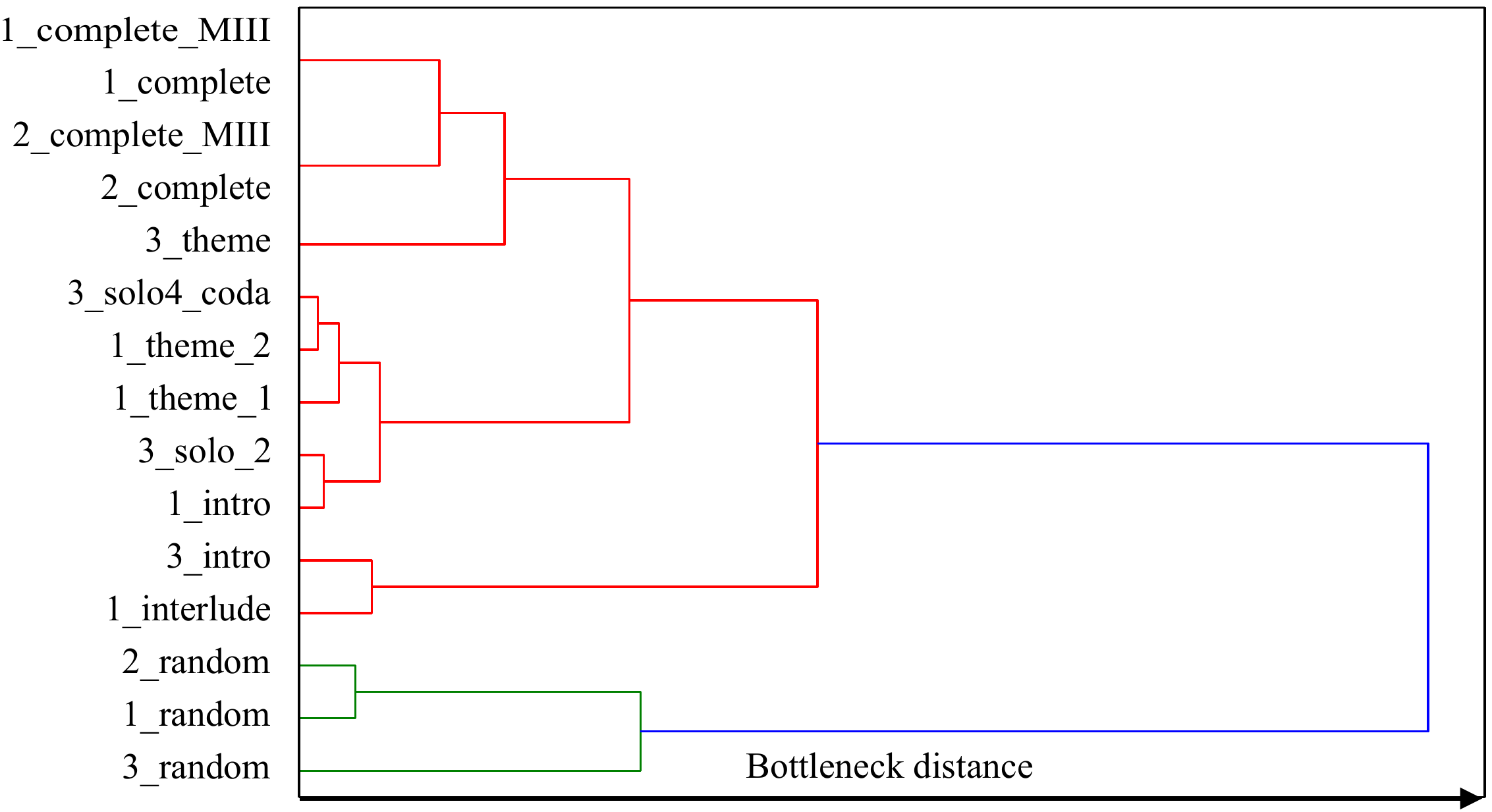}
\par\end{centering}
\protect\caption{Comparing three different versions of All the Things You Are.\label{fig:dend0NH}}
\end{figure}
\newline\newline
\textbf{Comparing three versions of \emph{All the Things You Are}.}
The three interpretations are structured as follows:
\begin{enumerate}%[a)]
\item Version $1$ is played by a quartet in a
standard way. Its main sections are a $3/4$
introduction, a first exposition of the theme, and a $12$
bars interlude introducing the last
theme enriched by short improvisations.

\item Version $2$ is performed by a piano solo, and it is characterized
by a rich chromatic playing style of both hands. The main
theme is executed twice.

\item Version $3$ is performed by a duo (piano and
bass). Its structure consists of an introduction, an
exposition of the theme, and a piano improvisation.
\end{enumerate}

\noindent The dataset is composed by the complete versions of the standard labeled as $i\_complete$ and a transposed version ($i\_complete\_interval$). Segments of each versions are included in the dataset and labeled as $i\_segment$.
In order to test the ability of the model to
distinguish between a piece modulating in several
tonalities, enriched with
chromatic solos, and a non-structured sequence of pitches, a random version of each interpretation is also part of the dataset ($i\_random$).

The resulting dendrogram is displayed in~\Cref{fig:dend0NH}. We observe that the transposed versions have distance
zero from the original ones, as an effect of the invariance of the filtration induced on the \textit{Tonnetz} torus by the height function under uniform transposition. The randomized versions of the
songs are well segregated. A small
cluster groups the interlude of the first and the introduction of
the third version, because both fragments share a very similar structure
in terms of intervallic leaps and rhythm. Finally, in the top
cluster, the two complete songs are linked to the fragment of the
third version containing the theme. Hence, the $0$th persistence
homology retrieves the fragments containing the whole structure of
the standard. This feature is surprising when taking into account the
several modulations of the piece.
\begin{figure}[tb]
\begin{centering}
\includegraphics[width=0.63\textwidth]{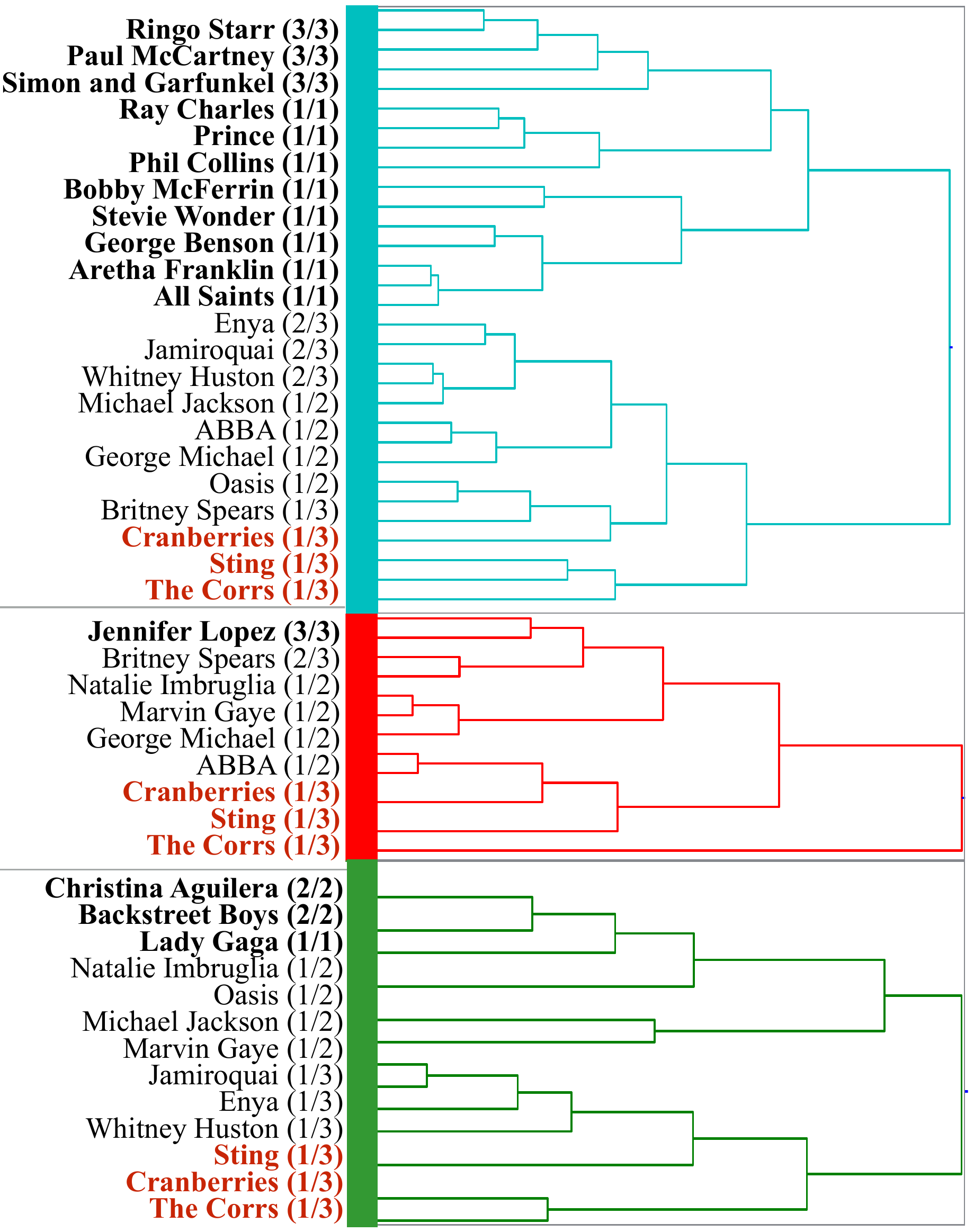}
\par\end{centering}
\protect\caption{A simplified representation of the clustering of $58$ pop songs generated from their $1$st persistence diagrams.\label{fig:bigpop}}
\end{figure}
\newline\newline
\textbf{Big Pop Clustering.} \Cref{fig:bigpop} shows a simplification of the clustering resulting by the comparison of the $1$st persistence diagrams associated with $58$ pop songs performed by $28$ artists, spacing from Ray Charles to Lady Gaga. In order to give a simplified representation of this dendrogram, we considered only the three biggest clusters detected by the algorithm. On the left of each cluster, we listed the artists whose songs belong to that group. Names written in black bold characters indicate artists whose songs are entirely grouped in the cluster at their right, while red bold characters identify the three artists whose songs are spread among the three groups. We observe how the entire collection of songs by {Ringo Starr}, {Paul McCartney} and {Simon \& Garfunkel} are grouped together in the blue cluster with {Ray Charles}, {Stevie Wonder} and {George Benson}.  Moreover, the heterogeneity that characterizes Sting's compositions is mirrored by the presence of one of his songs in each cluster. The second and third clusters are less homogeneous, but promising, taking into account that so far each song is identified by a single persistence diagram.

\section{Discussion and future works}

We suggested a model describing music by taking into account the
contribution of each pair (pitch class, duration) associated with
the notes of a composition. The height function has been defined on the vertices the simplicial \textit{Tonnetz} to induce a lower level set  filtration on the \textit{Tonnetz} torus. The $0$th and $1$st persistence diagrams associated with different musical pieces
have been interpreted in musical terms and their bottleneck distance has been used to classify them hierarchically. The possible
clusterings have been
represented as dendrograms, showing that $0$th and $1$st persistence
can be used to analyze and classify music.

%\begin{figure}[tb]
%\centering
%\includegraphics[width=0.8\textwidth]{img/chroma2.pdf}
%\caption{Chromagrams. A perfect cadence $Dm7-G7-Cmaj9-Cmaj7$ played by a Fender Rhodes (left) and a piano (right).\label{fig:chromas}}
%\end{figure}

The analysis and classification of music we performed has been
realized by considering datasets composed by MIDI files. However, the extension of this model to audio files is straightforward. Given an audio signal, the
chroma analysis~\cite{harte2005automatic} retrieves the contribution in time of each pitch class. Using a chromagram to define the height function, it would surely be affected by the noisy data coming
form the signal. However, the stability of the persistence
diagrams, when compared using the bottleneck distance, assures that a small perturbation of
the function inducing the filtration corresponds to small
variations of the persistence diagrams.

The model itself can be extended in several ways. For instance, it is possible
to augment the dimensionality the simplicial \textit{Tonnetz}. This would result in losing its property to be easily
visualizable, but it would give the possibility to encode more
information. This could be done by associating with
each pitch class of the \textit{Tonnetz} a velocity, or by adding
information concerning whatever pitch-class related feature. Moreover, topological persistence offers further tools
to improve the strategies we suggested. A natural development is
the study of the multidimensional persistent homology
\cite{carlsson2009theory,cerri2013betti} of
musical spaces and their time-varying nature~\cite{Bergomi2015dynamical}.

{\small
\bibliographystyle{splncs03}
\bibliography{RefsCTIC}
}
\end{document}